\documentclass{article}
\usepackage{spconf,amsmath,epsfig}
\usepackage{array}
\usepackage{multirow}
\usepackage{epstopdf}
\usepackage{booktabs}
\usepackage{subfigure}
\usepackage[numbers,sort&compress]{natbib}
\usepackage{algpseudocode}
\usepackage{setspace}
\usepackage{xcolor}
\usepackage{marvosym}
\let\OLDthebibliography\thebibliography
\renewcommand\thebibliography[1]{
  \OLDthebibliography{#1}
  \setlength{\parskip}{0pt}
  \setlength{\itemsep}{0pt plus 0.3ex}
}

\begin{document}\sloppy

\def\x{{\mathbf x}}
\def\L{{\cal L}}

\title{Deep Bi-directional Attention Network for Image Super-Resolution Quality Assessment}
%
\name{Yixiao Li$^{\star}$ \qquad Xiaoyuan Yang$^{\star}$\textsuperscript{\Letter} \qquad Jun Fu${^{\star}}^{\star}$ \qquad Guanghui Yue$^{\dagger}$ \qquad Wei Zhou$^{\ddagger}$\textsuperscript{\Letter}}
\address{$^{\star}$ School of Mathematical Sciences, Beihang University, China\\
${^{\star}}^{\star}$ School of Information Sciences, University of Science and Technology of China\\
$^{\dagger}$ School of Biomedical Engineering, Shenzhen University, China\\
$^{\ddagger}$ School of Computer Science and Informatics, Cardiff University, United Kingdom\\
Email: 18335310648@163.com, xiaoyuanyang@vip.163.com \qquad zhouw26@cardiff.ac.uk}


\maketitle

\begin{abstract}
There has emerged a growing interest in exploring efficient quality assessment algorithms for image super-resolution (SR). However, employing deep learning techniques, especially dual-branch algorithms, to automatically evaluate the visual quality of SR images remains challenging. Existing SR image quality assessment (IQA) metrics based on two-stream networks lack interactions between branches. To address this, we propose a novel full-reference IQA (FR-IQA) method for SR images. Specifically, producing SR images and evaluating how close the SR images are to the corresponding HR references are separate processes. Based on this consideration, we construct a deep \textbf{Bi-directional Attention Network} (BiAtten-Net) that dynamically deepens visual attention to distortions in both processes, which aligns well with the human visual system (HVS). Experiments on public SR quality databases demonstrate the superiority of our proposed BiAtten-Net over state-of-the-art quality assessment methods. In addition, the visualization results and ablation study show the effectiveness of bi-directional attention.
\end{abstract}
\section{Introduction}
\label{sec:intro}

Image super-resolution (SR) aims to reconstruct SR images from the input low-resolution (LR) images. Early image SR methods were based on interpolation, such as bicubic, cubic spline interpolation~\cite{1163711}, and adaptive structure kernels~\cite{951537}. Later, compressed sensing was applied to image SR, leading to many SR algorithms on the basis of sparse coding~\cite{5466111}. Recently, with the rapid development of deep learning, there has been a shift towards designing deep learning based image SR frameworks, including convolutional neural network (CNN) based methods~\cite{SRdensenet}, generative adversarial neural network (GAN) based methods~\cite{esrgan}, and attention based models~\cite{LIU20226179}. However, image SR is highly ill-posed, as the input LR image can be zoomed into SR results with various quality scores by different SR algorithms and upsampling factors. Therefore, the accurate quality evaluation of SR images is a vital but challenging problem.

In general, image quality assessment (IQA) methods can be categorized into full-reference (FR), reduced-reference (RR), and no-reference (NR) IQA methods according to the presence or full/partial absence of reference information. The simplest FR-IQA method is peak signal-to-noise ratio (PSNR) or mean square error (MSE). However, they only compute pixel differences by signal fidelity. Based on the characteristics of the human visual system (HVS), Wang et al. proposed the structural similarity (SSIM)~\cite{ssim}, which serves as the basis for many other metrics. These include the multi-scale SSIM (MS-SSIM)~\cite{msssim}, the complex wavelet SSIM (CW-SSIM)~\cite{cwssim}, the gradient magnitude similarity deviation (GMSD)~\cite{gmsd}, IQA using Kernel Sparse Coding (KSCM)~\cite{KCSM}, etc. However, these classical IQA methods are all designed for natural images instead of the SR scenario.

\begin{figure*}[t]
    \centering
    \includegraphics[scale=0.44]{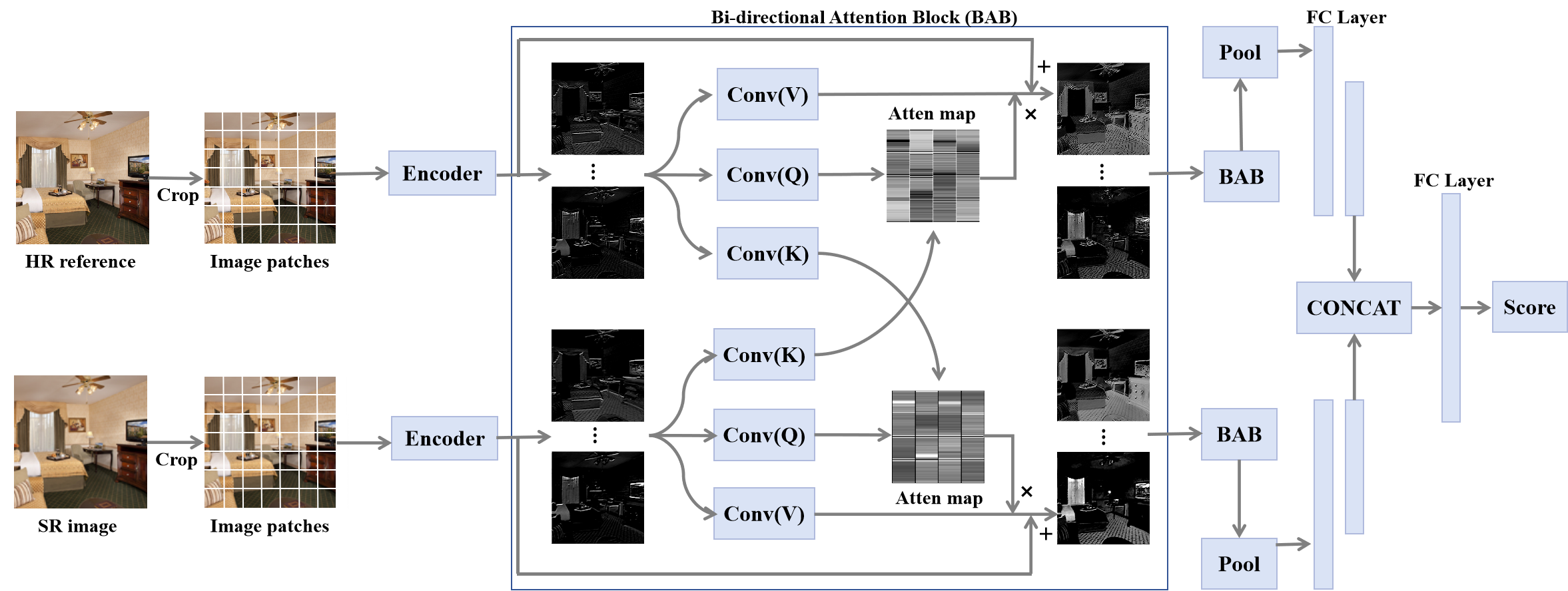}
    \caption{The overall framework of the proposed BiAtten-Net. \textbf{Encoder} indicates the stacking of a convolution layer, a batch normalization layer, and the activation function (ReLU). \textbf{Conv} refers to convolution layer, and the kernel size of all Convs is $3\times 3$. \textbf{Pool} represents the adaptive average pooling layer, and \textbf{Atten map} is the attention map. \textbf{$+$} refers to Shortcut, which directly adds the inputs of BAB to the end of the block. }
    \label{1}
\end{figure*}

Since SR images meet unique distortions that are different from conventional natural images, some hand-crafted FR-IQA methods developed for SR images have been proposed. To be specific, these methods consist of the structure-texture decomposition based algorithm, namely SIS~\cite{sis}, and the structural fidelity versus statistical naturalness (SFSN) method~\cite{sfsn}, as well as the deterministic and statistical fidelity method called SRIF ~\cite{SRIF}. Basically, they analyze the statistical characteristics of various SR images from a separate two-dimensional perspective. Although the predicted results of both dimensions are finally combined by weighting strategies, they generally lack the information interactions between the quality dimensions. Besides, deep learning based IQA models for SR images are mostly designed as NR-IQA methods. DeepSRQ~\cite{deepsqr} built a two-stream CNN to obtain the structural and texture features separately. HLSRIQA~\cite{zhang} developed a deep learning based NR-IQA method by the high-frequency and low-frequency maps of SR images. EK-SR-IQA~\cite{eksriqa} predicted SR image quality by leveraging a semi-supervised knowledge distillation strategy. 

On the contrary, existing deep learning based FR-IQA methods are designed for general image distortions on traditional IQA databases (e.g., LIVE~\cite{LIVE} and TID2013~\cite{TID2013}) rather than specific SR artifacts. For example, MGCN~\cite{MGCN} proposed the mask gated convolutional network for evaluating the image quality and identifying distortions simultaneously. WaDIQaM~\cite{bosse} proposed joint learning of local quality and local weights. LPIPS~\cite{LPIPS} calculated the distance of features extracted from the pre-trained networks between the reference and distorted images. AHIQ~\cite{Attenfr} utilized a two-stream network to extract the feature from both vision transformer~\cite{vit} and CNN branches. However, to the best of our knowledge, there are no deep learning based FR-IQA methods focusing on SR images.

Given that paying visual attention to artifacts of SR images aligns with the HVS, many works enhanced IQA's ability to capture quality degradations of local artifacts or dominant distorted regions by weighting attention maps. For example, JCSAN~\cite{jcsan} proposed a dual-branch based network to capture perceptual distortions based on joint channel-spatial attention. TADSRNet~\cite{tadsrnet} constructed a triple attention mechanism to acquire more significant portions of SR images. However, these attention-based methods lack interactions between individual branches (i.e., channel and spatial). Additionally, these methods only consider visual attention towards SR images and lack interactions with HR references.

To address the above-mentioned problems, we propose a deep Bi-directional Attention Network (BiAtten-Net). The main contributions of this work are summarized as follows:

1) Motivated by the properties of the HVS, we introduce the first deep learning based FR-IQA method (i.e., BiAtten-Net), which is specifically designed for SR images.

2) We propose a bi-directional attention mechanism that can dynamically simulate the processes of SR images approximating HR references and vice versa. This approach directly provides visual attention to distortions, thereby predicting quality scores that are more in line with the HVS.

3) Our method outperforms state-of-the-arts, especially achieving significant improvements over other FR-IQA methods regarding both natural and SR scenarios. Additionally, the visualization results and ablation study demonstrate the importance of bi-directional attention.

\begin{figure*}[t]
\centering
\includegraphics[scale=0.48]{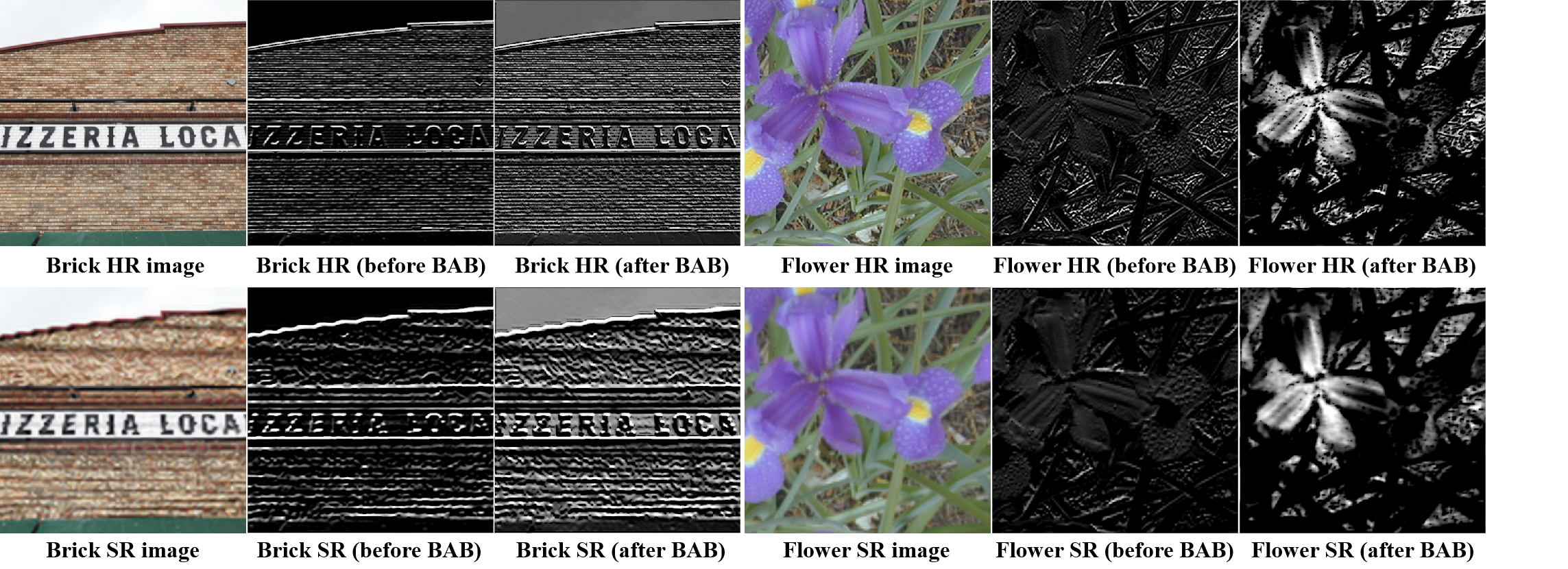}
    \caption{Visualization comparisons of feature maps regarding the proposed bi-directional attention block. \textbf{Brick HR image} and \textbf{Flower HR image} are HR references. \textbf{Brick SR image} and \textbf{Flower SR image} are SR images. The remaining images are feature maps before and after BAB in two branches.}
    \label{2}
\end{figure*}

\section{Proposed Method}
Current dual-branch based IQA methods for SR images lack interactions between sub-branches (e.g., structure and texture). Considering that the distortion arises from generating SR images from downsampled HR references, we dynamically simulate this process by approximating HR references to SR images in the branch with the HR reference as input. Furthermore, human subjects evaluate the perceptual quality of SR images by assessing the level to which the SR images approximate HR references. Therefore, we use the branch with the SR image as input for approaching SR images to HR references, which simulates the process of subjective quality assessment and thus is more consistent with the HVS. In this way, we effectively enhance visual attention to distortions by dynamically imitating the interactions of transforming HR references into SR images and vice versa.

Recently, the attention mechanism has been widely adopted in Transformer models~\cite{vit} as follows: 
\begin{equation}
\small
\begin{aligned}
&Q=XW^{Q}, \\
&K=XW^{K}, \\
&V=XW^{V}, \\
&D=Var(Q K^{T}). 
\end{aligned}
\label{eq1}
\end{equation}
Due to the attention mechanism being obtained by calculating the dot product between $Q$, $K$, and $V$, all three matrices need to be square matrices. Given an input image $X$ of shape $M\times M$, we can obtain $Q (Query) \in R^{M\times M}$, $K (Key) \in R^{M\times M}$, and $V (Value) \in R^{M\times M}$ matrices through linear transformations $W_{Q}$, $W_{K}$, and $W_{V}$. The linear transformations are typically fully connected linear layers. $D$ is the variance of the dot product of $Q$ and $K$. Afterward, the attention score is computed by taking the dot product of $Q$ and $K$, followed by normalization using standard deviation and the Softmax function:
\begin{equation}
\small
\begin{aligned}
&\operatorname{Attention}(Q, K, V)=\operatorname{Softmax}\left(\frac{Q K^{\top}}{\sqrt{D}}\right) V.
\end{aligned}
\label{eq2}
\end{equation}
Here, the attention score reflects the similarity between each pixel and the other pixels in the image $X$, thereby achieving effective visual attention for the image. Inspired by the attention mechanism, we develop our BiAtten-Net, as shown in Fig. \ref{1}. A two-stream network is exploited to extract features from HR reference and SR image, while gradually enhancing information interactions and directing visual attention to distortions through the proposed Bi-directional Attention Block (BAB). These features are then combined from both branches to predict the final visual quality score. 

Specifically, given a pair of input images (i.e., SR image and the corresponding HR reference), we first crop the images into overlapping patches. Following the settings of~\cite{deepsqr}, the patch size is set to $32\times32$ in our experiments. We then employ a stacking of single convolution layer with batch normalization layer (BN) and activation function (ReLU) to preliminarily extract the features of image patches. After that, the feature maps of these image patches are fed into the proposed BAB. We calculate the $Q$, $K$, and $V$ matrices for both branches separately:
\begin{equation}
\small
\begin{aligned}
&{Q^{HR},K^{HR},V^{HR}}=Conv(X^{HR}), \\
&{Q^{SR},K^{SR},V^{SR}}=Conv(X^{SR}),\\
&D^{HR}=Var(Q^{HR} {(K^{SR})}^{T}), \\
&D^{SR}=Var(Q^{SR} {(K^{HR})}^{T}),
\end{aligned}
\label{eq3}
\end{equation}
where $X^{SR}\in R^{M\times M}$ and $X^{HR}\in R^{M\times M}$ represent the input feature maps to BAB. $D^{HR}$ and $D^{SR}$ are the variance of the dot product of $Q^{HR},K^{SR}$ and $Q^{SR}, K^{HR}$, respectively. Considering that the learning capacity of linear layers is limited, we employ convolutional layers rather than linear layers to obtain the  $Q$, $K$, and $V$ matrices. The $K^{HR}$ and $K^{SR}$ matrices of the two branches are exchanged to calculate the attention maps as:
\begin{equation}
\small
\begin{aligned}
&\operatorname{Attention^{HR}}(Q^{HR}, K^{SR}, V^{HR})\\
&=\operatorname{Softmax}\left(\frac{Q^{HR} {K^{SR}}^{\top}}{\sqrt{D^{HR}}}\right) V^{HR},\\
&\operatorname{Attention^{SR}}(Q^{SR}, K^{HR}, V^{SR})\\
&=\operatorname{Softmax}\left(\frac{Q^{SR} {K^{HR}}^{\top}}{\sqrt{D^{SR}}}\right) V^{SR}.
\end{aligned}
\label{eq4}
\end{equation}
By swapping $K^{SR}$ with $K^{HR}$, the resulting attention maps calculate the pixel-level similarity between the current branch features and the features from the other branch. In this way, the attention maps directly deepen the visual attention on subtle differing pixels (i.e., distortions). Taking into account that the Identity Shortcut~\cite{resnet} has been proven to effectively alleviate model overfitting issues, we use the Shortcut as the main architecture. After passing through two BABs, the feature maps of both branches are flattened and concatenated through fully connected layers to enhance information interactions, ultimately obtaining perceptual quality predictions.

To intuitively demonstrate that the proposed method effectively enhances the visual attention to distortions, we visualize the intermediate feature maps of the two branches separately, as illustrated in Fig. \ref{2}. It can be observed that the distortions between HR references and SR images are concentrated in the appearances of bricks in “Brick SR image” and flowers in “Flower SR image”. Compared to the representations of distortions in feature maps before BAB, the distorted patterns after BAB are noticeably clearer. Moreover, the details of distortions are significantly increased, indicating that our proposed bi-directional attention can effectively enhance the visual attention to distortions.

\begin{figure}[t]
\centering   \includegraphics[scale=0.70]{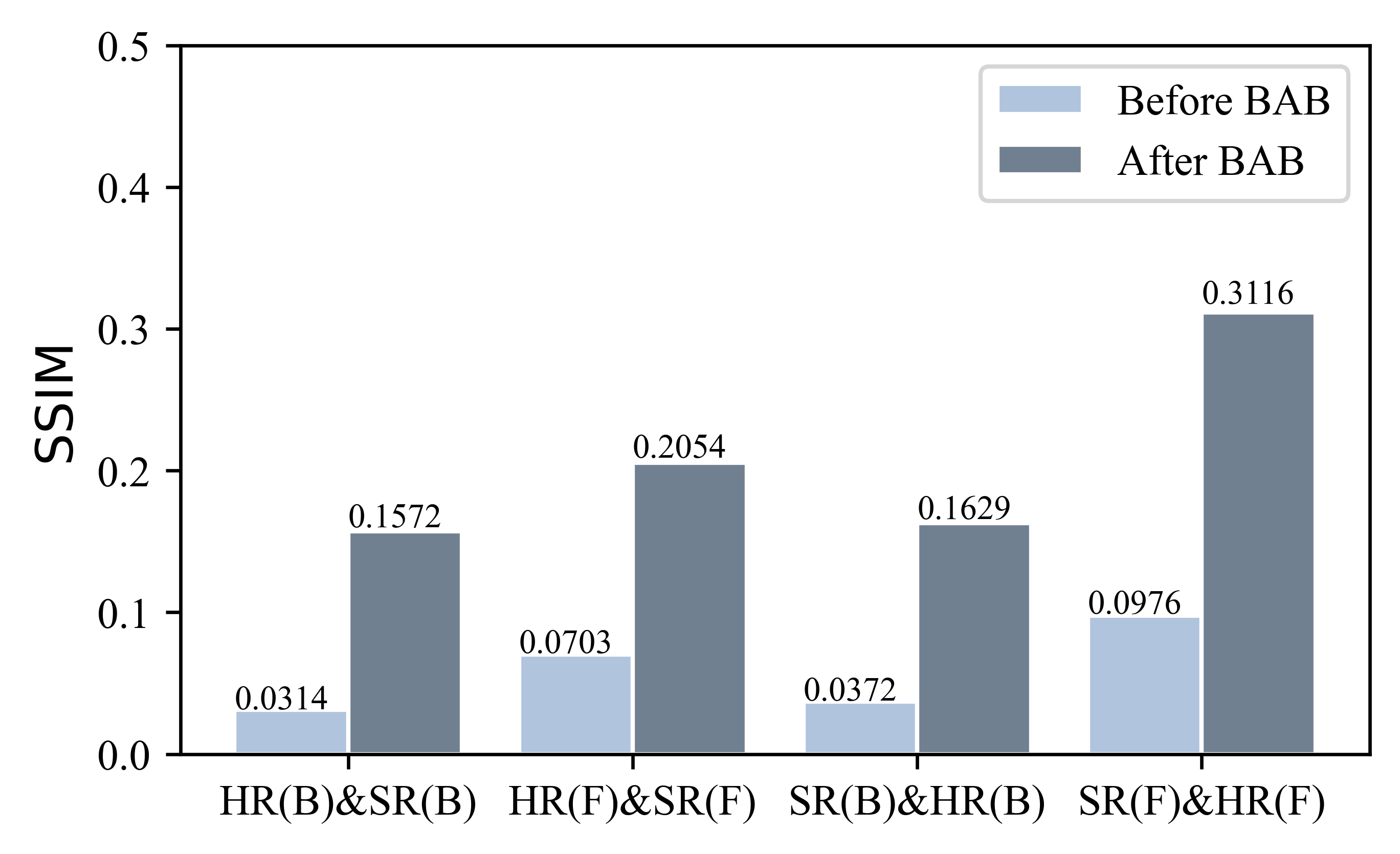}
    \caption{Similarity of “$X_{1}\& X_{2}$”, where $X_{1}$ is initial HR reference or SR image, $X_{2}$ is the feature map before or after BAB. $(B)$ and $(F)$ denote “Brick” and “Flower”, respectively. The higher the SSIM, the more similar the two images are.}
    \label{3}
\end{figure}

It is interesting to further illustrate the effectiveness of BAB. 
As mentioned earlier, the two branches simulate the processes of generating and evaluating distortions in SR images respectively. Therefore, in an ideal scenario, by continuous information interactions between branches, the feature maps of HR references can gradually approximate SR images, and vice versa. During the iterative learning, if the feature maps of SR images and HR references show significant improvement regarding similarity computation, the network can effectively simulate the processes of distortion generation and quality assessment, enabling a more comprehensive assessment of distortion level. Here, we calculate the similarity (e.g., SSIM) between the feature maps of the two branches before and after BAB, as shown in Fig. \ref{3}. The SSIM is used to reflect the level of approximation. It can be seen that the feature maps of the SR image are significantly improved in SSIM with the HR reference after getting visual attention from the BAB. Additionally, the feature maps of the HR reference also exhibit a significant SSIM improvement with the SR image after the BAB. This indicates our method dynamically pays visual attention to distortions as the “HR” and “SR” transform into each other.

\begin{table*}[t]
\small
\renewcommand{\arraystretch}{0.5}
  \centering
  \caption{Performance comparisons on QADS ~\cite{sis} and CVIU ~\cite{cviu} quality databases, where the best performance values of FR and NR are in \textcolor{red}{red} and \textcolor{blue}{blue}, respectively.}
    \begin{tabular}{c|c|cccc|cccc}
    \toprule
          &       & \multicolumn{4}{c|}{QADS}     & \multicolumn{4}{c}{CVIU} \\
    \midrule
    Types & Methods & SRCC  & KRCC  & PLCC  & RMSE  & SRCC  & KRCC  & PLCC  & RMSE \\
    \midrule
    \multirow{8}[2]{*}{FR-IQA} & PSNR  & 0.354 & 0.244 & 0.390 & 0.253 & 0.566 & 0.394 & 0.578 & 1.962 \\
          & SSIM~\cite{ssim}  & 0.529 & 0.369 & 0.533 & 0.233 & 0.629 & 0.443 & 0.650 & 1.828 \\
          & MS-SSIM~\cite{msssim} & 0.717 & 0.530 & 0.724 & 0.190 & 0.805 & 0.601 & 0.811 & 1.405 \\
          & CW-SSIM~\cite{cwssim} & 0.326 & 0.228 & 0.379 & 0.254 & 0.759 & 0.541 & 0.754 & 1.579 \\
          & GMSD~\cite{gmsd}  & 0.765 & 0.569 & 0.775 & 0.174 & 0.847 & 0.650 & 0.850 & 1.267 \\
          & WaDIQaM~\cite{bosse} & 0.871 & \_ & 0.887 & 0.128 & 0.872 & \_ & 0.886 & 1.304 \\
          & LPIPS ~\cite{LPIPS}& 0.881 & \_ & 0.873 & 0.129 & 0.849 & \_ & 0.852 & 1.313 \\
    \midrule
    \multirow{3}[2]{*}{NR-IQA} & NIQE~\cite{niqe}  & 0.398 & 0.279 & 0.404 & 0.251 & 0.653 & 0.478 & 0.666 & 1.794 \\
          & LPSI~\cite{lpsi}  & 0.408 & 0.289 & 0.422 & 0.249 & 0.488 & 0.350 & 0.537 & 2.027 \\
          & MetaIQA ~\cite{metaiqa}& 0.826 & \_ & 0.790 & 0.178 & 0.720 & \_ & 0.746 & 1.718 \\
          & HyperIQA~\cite{hyperiqa} & 0.954 & 0.815 & 0.957 & 0.099 & 0.933 & 0.772 & 0.928 & 1.017 \\
    \midrule
    \multirow{5}[2]{*}{SR NR-IQA} & DeepSRQ~\cite{deepsqr} & 0.953 & \_    & 0.956 & 0.077 & 0.921 & \_    & 0.927 & 0.904 \\
          & HLSRIQA~\cite{zhang} & 0.961 & 0.829 & 0.950 & 0.741 & 0.948 & 0.810 & 0.948 & 0.775 \\
          & EK-SR-IQA~\cite{eksriqa} & 0.963 & \_ & 0.966 & \_ & 0.953 & \_ & 0.951 & \_ \\
          & JCSAN~\cite{jcsan} & 0.971 & 0.858 & 0.973 & \textcolor{blue}{0.065} & 0.949 & 0.808 & 0.957 & 0.777\\
          & TADSRNet~\cite{tadsrnet} & \textcolor{blue}{0.972} & \textcolor{blue}{0.862} & \textcolor{blue}{0.974} & 0.067 & \textcolor{blue}{0.952} & \textcolor{blue}{0.812} & \textcolor{blue}{0.959} & \textcolor{blue}{0.797}\\
          \midrule
    \multirow{4}[2]{*}{SR FR-IQA} & SIS~\cite{sis} & 0.913 & 0.740 & 0.914 & 0.112 & 0.869 & 0.686 & 0.897 & 1.061 \\
          & SFSN~\cite{sfsn} & 0.841 & 0.655 & 0.845 & 0.147 & 0.871 & 0.680 & 0.885 & 1.120 \\
          & SRIF~\cite{SRIF} & 0.916 & 0.746 & 0.917 & 0.109 & 0.886 & 0.704 & 0.902 & 1.039 \\
          & Proposed BiAtten-Net & \textcolor{red}{0.981} & \textcolor{red}{0.895} & \textcolor{red}{0.982} & \textcolor{red}{0.055} & \textcolor{red}{0.972} & \textcolor{red}{0.862} & \textcolor{red}{0.976} & \textcolor{red}{0.515} \\
    \bottomrule
    \end{tabular}%
  \label{t2}%
\end{table*}%

\section{Validation}

\subsection{Experimental Protocols}
We conduct experiments on QADS~\cite{sis} and CVIU~\cite{cviu} databases. The QADS database contains 20 original HR references and 980 SR images created by 21 SR algorithms, including 4 interpolation-based, 11 dictionary-based, and 6 DNN-based SR models, with upsampling factors equaling 2, 3, and 4. Each SR
image is associated with the mean opinion score (MOS) of 100 subjects. In the CVIU database, 1620 SR images are produced by 9 SR approaches from 30 HR references. Six pairs of scaling factors and kernel widths are adopted, where a larger subsampling factor corresponds to a larger blur kernel width. Each image is rated by 50 subjects, and the mean of the median 40 scores is calculated for each image as the MOS.


The QADS and CVIU databases are randomly divided into non-overlapping 80$\%$ and 20$\%$ sets, with $80\%$ of the data used for training and the remaining $20\%$ for testing. We train on QADS and CVIU training sets for 500 epochs and 300 epochs, respectively. These epochs were chosen based on experimental experience to ensure sufficient convergence of the network while avoiding overfitting caused by excessive training. We use L1-loss to measure the difference between predicted scores and MOSs. The optimizer used is stochastic gradient descent (SGD), with an initial learning rate of $0.01$, momentum of $0.9$, and weight decay setting to $10^{-6}$.

We adopt four commonly used evaluation criteria to compare performance, including Spearman rank-order correlation coefficient (SRCC), Kendall rank-order correlation coefficient (KRCC), Pearson linear correlation coefficient (PLCC), and root mean square error (RMSE). SRCC and PLCC/RMSE are employed to assess the monotonicity and accuracy of predictions, respectively. KRCC is used to measure the ordinal association between two measured quantities. An ideal quality metric would have SRCC, KRCC, and PLCC values close to one, and RMSE close to zero. It should be noted that a five-parameter nonlinear fitting process~\cite{SRIF} is applied to map the predicted qualities into a standardized scale of subjective quality labels before calculating PLCC and RMSE.

\subsection{Performance Comparisons}
To validate the proposed method, we compare it with state-of-the-art FR-IQA, NR-IQA, and SR IQA methods. FR-IQA methods include PSNR, SSIM~\cite{ssim}, MS-SSIM~\cite{msssim}, CW-SSIM~\cite{cwssim}, GMSD~\cite{gmsd}, WaDIQaM~\cite{bosse}, and LPIPS~\cite{LPIPS}. NR-IQA methods consist of the NIQE~\cite{niqe}, LPSI~\cite{lpsi}, MetaIQA~\cite{metaiqa}, and HyperIQA~\cite{hyperiqa}. Among them, WaDIQaM, LPIPS, MetaIQA, and HyperIQA are deep learning based models. SR IQA methods contain SIS~\cite{sis}, SFSN~\cite{sfsn}, SRIF~\cite{SRIF}, DeepSRQ~\cite{deepsqr}, HLSRIQA~\cite{zhang}, EK-SR-IQA~\cite{eksriqa}, JCSAN~\cite{jcsan}, and TADSRNet~\cite{tadsrnet}. Besides, DeepSRQ, HLSRIQA, EK-SR-IQA, JCSAN and TADSRNet are deep learning based methods.


The comparison results are shown in Table \ref{t2}. In general, deep learning based methods have better performance, and the performance of FR-IQA and NR-IQA methods is basically inferior to SR IQA methods, indicating that traditional IQA methods cannot cover diverse artifacts of SR images. Among SR IQA methods, FR methods (i.e., SIS, SFSN, SRIF) are limited to shallow features and cannot fully utilize the hidden information in reference images, causing significant gaps compared to those deep learning based methods. In addition, our proposed BiAtten-Net achieves greater information interactions between branches and directly pays visual attention to distortions, which effectively exploits the deep features of the reference image. Therefore, the proposed method achieves the best performance on both QADS and CVIU databases.

\subsection{Ablation Study}
To verify the effectiveness of bidirectional information interactions in BAB, we conduct an ablation study on the interactive modes of visual attention. Specifically, we validate the following scenarios: without using BAB (i.e., separately applying the attention in Eq. \ref{eq1} to SR image and HR reference branches); adding the attention information from the SR image branch to the HR reference branch while applying the attention in Eq. \ref{eq1} to the SR image branch (i.e., SR → HR); and adding the attention information from the HR reference branch to the SR image branch while applying the attention in Eq. \ref{eq1} to the HR reference branch (i.e., HR → SR). The experimental results find out that the model without BAB has the worst performance on both databases, and even adding one-way attention information interaction can significantly improve performance. In addition, the performance improvement brought by BAB is particularly significant in KRCC and RMSE on both databases. Ultimately, the model using BAB achieves the best performance, indicating that our proposed BAB effectively enhances the network's learning ability.
\begin{table}[t]
\renewcommand{\arraystretch}{0.8}
  \centering
  \caption{Ablation study of bi-directional attention. HR → SR and SR → HR represent one-way attentional information interaction, where the attention maps for both branches are calculated using $K$ matrix from the other branch.}
  \resizebox{\linewidth}{!}{
    \begin{tabular}{c|cccc|cccc}
    \toprule
          & \multicolumn{4}{c|}{QADS}     & \multicolumn{4}{c}{CVIU} \\
    \midrule
    Models & SRCC  & KRCC  & PLCC  & RMSE  & SRCC  & KRCC  & PLCC  & RMSE \\
    \midrule
    w/o BAB &0.936 & 0.795 & 0.937 & 0.103 & 0.942 & 0.808 & 0.954 & 0.815 \\
            HR→SR & 0.957 & 0.840 & 0.958 & 0.079 & 0.970 & 0.853 & 0.971 & 0.706  \\
           SR→HR & 0.955 & 0.834 & 0.957 & 0.082 & 0.960 & 0.829 & 0.960 & 0.763 \\
           with BAB & 0.981 & 0.895 & 0.982 & 0.055 & 0.972 & 0.862 & 0.976 & 0.515 \\ 
    \bottomrule
    \end{tabular}}
  \label{t3}%
\end{table}%
\section{Conclusion}
In this paper, we are the first to propose a deep learning based FR-IQA method specifically designed for SR images. Inspired by the characteristics of the HVS, we introduce bi-directional attention tailored for SR images. This pioneers a new model for learning distortions in SR images through the interactions of bi-directional information between SR images and the corresponding HR references. Experimental results demonstrate that our proposed BiAtten-Net effectively provides visual attention to SR distortions and surpasses existing state-of-the-art quality assessment methods. The codes are publicly available at https://github.com/Lighting-YXLI/BiAtten-Net.

\small
\bibliographystyle{IEEEbib}
\bibliography{icme2023template}

\begin{thebibliography}{10}

\bibitem{1163711}
R.~Keys,
\newblock ``Cubic convolution interpolation for digital image processing,''
\newblock {\em IEEE Transactions on Acoustics, Speech, and Signal Processing},
  vol. 29, no. 6, pp. 1153--1160, 1981.

\bibitem{951537}
Xin Li and M.T. Orchard,
\newblock ``New edge-directed interpolation,''
\newblock {\em IEEE TIP}, vol. 10, no. 10, pp. 1521--1527, 2001.

\bibitem{1658087}
Lei Zhang and Xiaolin Wu,
\newblock ``An edge-guided image interpolation algorithm via directional
  filtering and data fusion,''
\newblock {\em IEEE TIP}, vol. 15, no. 8, pp. 2226--2238, 2006.

\bibitem{105555}
Philippe Th\'{e}venaz, Thierry Blu, and Michael Unser,
\newblock {\em Image Interpolation and Resampling}, p. 393–420,
\newblock Academic Press, Inc., 2000.

\bibitem{8336891}
Yunfeng Zhang, Qinglan Fan, Fangxun Bao, Yifang Liu, and Caiming Zhang,
\newblock ``Single-image super-resolution based on rational fractal
  interpolation,''
\newblock {\em IEEE TIP}, vol. 27, no. 8, pp. 3782--3797, 2018.

\bibitem{5466111}
Jianchao Yang, John Wright, Thomas~S. Huang, and Yi~Ma,
\newblock ``Image super-resolution via sparse representation,''
\newblock {\em IEEE TIP}, vol. 19, no. 11, pp. 2861--2873, 2010.

\bibitem{4587647}
Jianchao Yang, John Wright, Thomas Huang, and Yi~Ma,
\newblock ``Image super-resolution as sparse representation of raw image
  patches,''
\newblock in {\em CVPR}, 2008, pp. 1--8.

\bibitem{bosse}
Sebastian Bosse, Dominique Maniry, Klaus-Robert Müller, Thomas Wiegand, and
  Wojciech Samek,
\newblock ``Deep neural networks for no-reference and full-reference image
  quality assessment,''
\newblock {\em IEEE TIP}, vol. 27, no. 1, pp. 206--219, 2018.

\bibitem{LPIPS}
Richard Zhang, Phillip Isola, Alexei~A. Efros, Eli Shechtman, and Oliver Wang,
\newblock ``The unreasonable effectiveness of deep features as a perceptual
  metric,''
\newblock in {\em CVPR}, 2018.

\bibitem{metaiqa}
Hancheng Zhu, Leida Li, Jinjian Wu, Weisheng Dong, and Guangming Shi,
\newblock ``{MetaIQA}: Deep meta-learning for no-reference image quality
  assessment,''
\newblock in {\em CVPR}, 2020.

\bibitem{hyperiqa}
Shaolin Su, Qingsen Yan, Yu~Zhu, Cheng Zhang, Xin Ge, Jinqiu Sun, and Yanning
  Zhang,
\newblock ``Blindly assess image quality in the wild guided by a self-adaptive
  hyper network,''
\newblock in {\em CVPR}, 2020.

\bibitem{srcnn}
Chao Dong, Chen~Change Loy, Kaiming He, and Xiaoou Tang,
\newblock ``Image super-resolution using deep convolutional networks,''
\newblock {\em IEEE PAMI}, vol. 38, no. 2, pp. 295--307, 2016.

\bibitem{SRdensenet}
Tong Tong, Gen Li, Xiejie Liu, and Qinquan Gao,
\newblock ``Image super-resolution using dense skip connections,''
\newblock in {\em ICCV}, 2017.

\bibitem{srgan}
Christian Ledig, Lucas Theis, Ferenc Huszár, Jose Caballero, Andrew
  Cunningham, Alejandro Acosta, Andrew Aitken, Alykhan Tejani, Johannes Totz,
  Zehan Wang, and Wenzhe Shi,
\newblock ``Photo-realistic single image super-resolution using a generative
  adversarial network,''
\newblock in {\em CVPR}, 2017, pp. 105--114.

\bibitem{esrgan}
Xintao Wang, Ke~Yu, Shixiang Wu, Jinjin Gu, Yihao Liu, Chao Dong, Yu~Qiao, and
  Chen~Change Loy,
\newblock ``{ESRGAN}: Enhanced super-resolution generative adversarial
  networks,''
\newblock in {\em ECCV}, 2019.

\bibitem{ssim}
Zhou Wang, A.C. Bovik, H.R. Sheikh, and E.P. Simoncelli,
\newblock ``Image quality assessment: from error visibility to structural
  similarity,''
\newblock {\em IEEE TIP}, vol. 13, no. 4, pp. 600--612, 2004.

\bibitem{ifc}
H.R. Sheikh, A.C. Bovik, and G.~de~Veciana,
\newblock ``An information fidelity criterion for image quality assessment
  using natural scene statistics,''
\newblock {\em IEEE TIP}, vol. 14, no. 12, pp. 2117--2128, 2005.

\bibitem{gmsd}
Wufeng Xue, Lei Zhang, Xuanqin Mou, and Alan~C. Bovik,
\newblock ``Gradient magnitude similarity deviation: A highly efficient
  perceptual image quality index,''
\newblock {\em IEEE TIP}, vol. 23, no. 2, pp. 684--695, 2014.

\bibitem{8461931}
Bahetiyaer Bare, Ke~Li, Bo~Yan, Bailan Feng, and Chunfeng Yao,
\newblock ``A deep learning based no-reference image quality assessment model
  for single-image super-resolution,''
\newblock in {\em ICASSP}, 2018, pp. 1223--1227.

\bibitem{Attenfr}
Shanshan Lao, Yuan Gong, Shuwei Shi, Sidi Yang, Tianhe Wu, Jiahao Wang, Weihao
  Xia, and Yujiu Yang,
\newblock ``Attentions help cnns see better: Attention-based hybrid image
  quality assessment network,''
\newblock in {\em CVPR Workshops}, 2022.

\bibitem{madnet}
Rushi Lan, Long Sun, Zhenbing Liu, Huimin Lu, Cheng Pang, and Xiaonan Luo,
\newblock ``Madnet: A fast and lightweight network for single-image super
  resolution,''
\newblock {\em IEEE Transactions on Cybernetics}, vol. 51, no. 3, pp.
  1443--1453, 2021.

\bibitem{LIU20226179}
Xiangbin Liu, Shuqi Chen, Liping Song, Marcin Woźniak, and Shuai Liu,
\newblock ``Self-attention negative feedback network for real-time image
  super-resolution,''
\newblock {\em Journal of King Saud University - Computer and Information
  Sciences}, vol. 34, no. 8, pp. 6179--6186, 2022.

\bibitem{cwssim}
Mehul~P. Sampat, Zhou Wang, Shalini Gupta, Alan~Conrad Bovik, and Mia~K.
  Markey,
\newblock ``Complex wavelet structural similarity: A new image similarity
  index,''
\newblock {\em IEEE TIP}, vol. 18, no. 11, pp. 2385--2401, 2009.

\bibitem{msssim}
Zhou Wang, E.P. Simoncelli, and A.C. Bovik,
\newblock ``Multiscale structural similarity for image quality assessment,''
\newblock in {\em ACSSC}, 2003.

\bibitem{fsim}
Lin Zhang, Lei Zhang, Xuanqin Mou, and David Zhang,
\newblock ``Fsim: A feature similarity index for image quality assessment,''
\newblock {\em IEEE TIP}, vol. 20, no. 8, pp. 2378--2386, 2011.

\bibitem{sis}
Fei Zhou, Rongguo Yao, Bozhi Liu, and Guoping Qiu,
\newblock ``Visual quality assessment for super-resolved images: Database and
  method,''
\newblock {\em IEEE TIP}, vol. 28, no. 7, pp. 3528--3541, 2019.

\bibitem{sfsn}
Wei Zhou, Zhou Wang, and Zhibo Chen,
\newblock ``Image super-resolution quality assessment: Structural fidelity
  versus statistical naturalness,''
\newblock in {\em QoMEX}, 2021.

\bibitem{SRIF}
Wei Zhou and Zhou Wang,
\newblock ``Quality assessment of image super-resolution: Balancing
  deterministic and statistical fidelity,''
\newblock in {\em ACM Multimedia}, 2022.

\bibitem{deepsqr}
Wei Zhou, Qiuping Jiang, Yuwang Wang, Zhibo Chen, and Weiping Li,
\newblock ``Blind quality assessment for image superresolution using deep
  two-stream convolutional networks,''
\newblock {\em Information Sciences}, vol. 528, pp. 205--218, 2020.

\bibitem{ZHOU2023117025}
Fei Zhou, Wei Sheng, Zitao Lu, Bo~Kang, Mianyi Chen, and Guoping Qiu,
\newblock ``Super-resolution image visual quality assessment based on
  structure–texture features,''
\newblock {\em Signal Processing: Image Communication}, vol. 117, pp. 117025,
  2023.

\bibitem{LIVE}
H.R. Sheikh, M.F. Sabir, and A.C. Bovik,
\newblock ``A statistical evaluation of recent full reference image quality
  assessment algorithms,''
\newblock {\em IEEE TIP}, vol. 15, no. 11, pp. 3440--3451, 2006.

\bibitem{TID2013}
Nikolay Ponomarenko, Oleg Ieremeiev, Vladimir Lukin, Karen Egiazarian, Lina
  Jin, Jaakko Astola, Benoit Vozel, Kacem Chehdi, Marco Carli, Federica
  Battisti, and C.-C.~Jay Kuo,
\newblock ``Color image database {TID}2013: Peculiarities and preliminary
  results,''
\newblock in {\em EUVIP}, 2013.

\bibitem{PIPAL}
Gu~Jinjin, Cai Haoming, Chen Haoyu, Ye~Xiaoxing, Jimmy~S. Ren, and Dong Chao,
\newblock ``{PIPAL}: A large-scale image quality assessment dataset for
  perceptual image restoration,''
\newblock in {\em ECCV}, 2020.

\bibitem{disq}
Tiesong Zhao, Yuting Lin, Yiwen Xu, Weiling Chen, and Zhou Wang,
\newblock ``Learning-based quality assessment for image super-resolution,''
\newblock {\em IEEE TMM}, vol. 24, pp. 3570--3581, 2021.

\bibitem{fei}
``Super-resolution image visual quality assessment based on structure–texture
  features,''
\newblock {\em Signal Processing: Image Communication}, vol. 117, pp. 117025,
  2023.

\bibitem{dist}
Keyan Ding, Kede Ma, Shiqi Wang, and Eero~P. Simoncelli,
\newblock ``Image quality assessment: Unifying structure and texture
  similarity,''
\newblock {\em IEEE TPAMI}, vol. 44, no. 5, pp. 2567--2581, 2022.

\bibitem{DBCNN}
Weixia Zhang, Kede Ma, Jia Yan, Dexiang Deng, and Zhou Wang,
\newblock ``Blind image quality assessment using a deep bilinear convolutional
  neural network,''
\newblock {\em IEEE TCSVT}, vol. 30, no. 1, pp. 36--47, 2020.

\bibitem{weizhang}
Wei Zhang, Ralph~R. Martin, and Hantao Liu,
\newblock ``A saliency dispersion measure for improving saliency-based image
  quality metrics,''
\newblock {\em IEEE TCSVT}, vol. 28, no. 6, pp. 1462--1466, 2018.

\bibitem{sdcnn}
Sen Jia and Yang Zhang,
\newblock ``Saliency-based deep convolutional neural network for no-reference
  image quality assessment,''
\newblock {\em Multimedia Tools and Applications}, vol. 77, pp. 14859--14872,
  2018.

\bibitem{jcsan}
Tingyue Zhang, Kaibing Zhang, Chuan Xiao, Zenggang Xiong, and Jian Lu,
\newblock ``Joint channel-spatial attention network for super-resolution image
  quality assessment,''
\newblock {\em Appl. Intell}, vol. 52, pp. 17118--17132, 2022.

\bibitem{tadsrnet}
Xing Quan, Kaibing Zhang, Hui Li, Dandan Fan, Yanting Hu, and Jinguang Chen,
\newblock ``{TADSRNet}: A triple-attention dual-scale residual network for
  super-resolution image quality assessment,''
\newblock {\em Appl. Intell}, vol. 53, pp. 26708–26724, 2023.

\bibitem{sdgnet}
Sheng Yang, Qiuping Jiang, Weisi Lin, and Yongtao Wang,
\newblock ``{SGDNet}: An end-to-end saliency-guided deep neural network for
  no-reference image quality assessment,''
\newblock in {\em ACM Multimedia}, 2019.

\bibitem{eksriqa}
Haiyu Zhang, Shaolin Su, Yu~Zhu, Jinqiu Sun, and Yanning Zhang,
\newblock ``Boosting no-reference super-resolution image quality assessment
  with knowledge distillation and extension,''
\newblock in {\em ICASSP}, 2023.

\bibitem{zhang}
Zicheng Zhang, Wei Sun, Xiongkuo Min, Wenhan Zhu, Tao Wang, Wei Lu, and
  Guangtao Zhai,
\newblock ``A no-reference deep learning quality assessment method for
  super-resolution images based on frequency maps,''
\newblock in {\em ISCAS}, 2022.

\bibitem{vit}
Alexey Dosovitskiy, Lucas Beyer, et~al.,
\newblock ``An image is worth 16x16 words: Transformers for image recognition
  at scale,''
\newblock {\em arXiv preprint arXiv:2010.11929}, 2020.

\bibitem{swin}
Ze~Liu, Yutong Lin, Yue Cao, Han Hu, Yixuan Wei, Zheng Zhang, Stephen Lin, and
  Baining Guo,
\newblock ``Swin transformer: Hierarchical vision transformer using shifted
  windows,''
\newblock in {\em ICCV}, 2021, pp. 9992--10002.

\bibitem{resnet}
Kaiming He, Xiangyu Zhang, Shaoqing Ren, and Jian Sun,
\newblock ``Deep residual learning for image recognition,''
\newblock in {\em CVPR}, 2016.

\bibitem{cviu}
Chao Ma, Chih-Yuan Yang, Xiaokang Yang, and Ming-Hsuan Yang,
\newblock ``Learning a no-reference quality metric for single-image
  super-resolution,''
\newblock {\em CVIU}, vol. 158, pp. 1--16, 2017.

\bibitem{brisque}
Anish Mittal, Anush~Krishna Moorthy, and Alan~Conrad Bovik,
\newblock ``No-reference image quality assessment in the spatial domain,''
\newblock {\em IEEE TIP}, vol. 21, no. 12, pp. 4695--4708, 2012.

\bibitem{niqe}
Anish Mittal, Rajiv Soundararajan, and Alan~C. Bovik,
\newblock ``Making a “completely blind” image quality analyzer,''
\newblock {\em IEEE Signal Processing Letters}, vol. 20, no. 3, pp. 209--212,
  2013.

\bibitem{lpsi}
Qingbo Wu, Zhou Wang, and Hongliang Li,
\newblock ``A highly efficient method for blind image quality assessment,''
\newblock in {\em ICIP}, 2015.

\bibitem{igts}
Lijuan Tang, Kezheng Sun, Luping Liu, Guangcheng Wang, and Yutao Liu,
\newblock ``A reduced-reference quality assessment metric for super-resolution
  reconstructed images with information gain and texture similarity,''
\newblock {\em Signal Processing: Image Communication}, vol. 79, pp. 32--39,
  2019.

\bibitem{MGCN}
Chen Huang, Tingting Jiang, and Ming Jiang,
\newblock ``Encoding distortions for multi-task full-reference image quality
  assessment,''
\newblock in {\em ICME}, 2019.

\bibitem{KCSM}
Zihan Zhou, Jing Li, Yuhui Quan, and Ruotao Xu,
\newblock ``Image quality assessment using kernel sparse coding,''
\newblock {\em IEEE TMM}, vol. 23, pp. 1592--1604, 2021.

\bibitem{pfsm}
Wenfei Wan, Jinjian Wu, Guangming Shi, Yongbo Li, and Weisheng Dong,
\newblock ``Super-resolution quality assessment: Subjective evaluation database
  and quality index based on perceptual structure measurement,''
\newblock in {\em ICME}, 2018.

\end{thebibliography}

\end{document}